\documentclass[12pt,tightenlines,eqsecnum,floats,aps,amsmath,amssymb,nofootinbib,prd,showpacs]{revtex4}

\usepackage{setspace} \usepackage{amsmath,amssymb,amsfonts,amsthm}
\usepackage{graphicx}

\newcommand{\be}{\begin{equation}}
\newcommand{\ee}{\end{equation}}
\newcommand{\beq}{\begin{eqnarray}}
\newcommand{\eeq}{\end{eqnarray}}

\def\lsim{\hbox{ \raise.35ex\rlap{$<$}\lower.6ex\hbox{$\sim$}\ }}
\def\gsim{\hbox{ \raise.35ex\rlap{$>$}\lower.6ex\hbox{$\sim$}\ }} 

\usepackage{bbold}

\begin{document}
\title{Numerical techniques for solving the quantum constraint equation of
  generic lattice-refined models in loop quantum cosmology} 
\author{William Nelson and Mairi Sakellariadou} 
\affiliation{King's College London, Department of Physics, Strand WC2R 
2LS, London, U.K.}

\begin{abstract}
\vspace{.2cm}
\noindent
To avoid instabilities in the continuum semi-classical limit of loop
quantum cosmology models, refinement of the underlying lattice is
necessary. The lattice refinement leads to new dynamical difference
equations which, in general, do not have a uniform step-size, implying
complications in their analysis and solutions. We propose a numerical
method based on Taylor expansions, which can give us the necessary
information to calculate the wave-function at any given lattice point.
The method we propose can be applied in any lattice-refined model,
while in addition the accuracy of the method can be estimated.
Moreover, we confirm numerically the stability criterion which was
earlier found following a von Neumann analysis. Finally, the `motion'
of the wave-function due to the underlying discreteness of the
space-time is investigated, for both a constant lattice, as well as
lattice refinement models.
\end{abstract}
 

\maketitle

\section{Introduction}

Loop quantum gravity~\cite{rovelli2004} canonically quantises
space-time via triads and holonomies of the Ashtekar
connection. Whilst a full understanding of the theory has yet to be
reached, symmetry reduced versions akin to the Wheeler-de Witt
mini-superspace model have been successfully
developed~\cite{Ashtekar:2003hd}.  As a first approximation the
quantised holonomies were taken to be shift operators {\it with a
  fixed magnitude}. This results in the quantised Hamiltonian
constraint being a difference equation with a constant interval
between points on the lattice. Whilst these models are reasonably
successful in studying certain aspects of the quantum
regime~\cite{Bojowald:2002gz,Ashtekar:2003hd}, it has been shown that
they lead to serious instabilities in the continuum, semi-classical
limit~\cite{Rosen:2006bga,Bojowald:2007ra}.  In the underlying loop
quantum gravity theory, the contributions to the (discrete)
Hamiltonian operator depend on the state which describes the
universe. As the volume grows (the universe expands), the number of
contributions increases. Thus, the Hamiltonian constraint operator
is expected to create new vertices of a lattice state (in addition
to changing their edge labels), which in loop quantum cosmology
result in a refinement of the discrete lattice.

It has been recently shown how this lattice refinement effect can be
modelled and how this approach eliminates the problematic
instabilities in the continuum era~\cite{Nelson:2007um}. Whilst the
continuum limit of these lattice refining models can be taken, there
is a complication in directly evolving two-dimensional wave-functions,
such as those necessary to study Bianchi models or black hole interiors.
The information needed to calculate the wave-function at a given lattice
point is not provided by previous iterations.  Recently, it has been
demonstrated~\cite{Sabharwal:2007xy} that a simple local interpolation
scheme can be used to approximate the necessary data points, allowing
direct numerical evolution of such two-dimensional systems.

In this paper, we show how Taylor expansions can be used to perform
this interpolation with a well-defined and predicable accuracy. We
first develop the scheme for the one-dimensional homogeneous,
isotropic cosmological case, which has analytic solutions. We use this
simple example to show that the accuracy of the system is well
controlled. We then study the two-dimensional case of a Schwarszchild
interior, which cannot be exactly solved (for a general lattice
refinement scheme).  As our interpolation system is based only on
Taylor expansions, all we require is that the function used to model
the lattice refinement be analytic. This allows us to look in detail
at the effects of general lattice refinement models, beyond the
simplest cases considered thus far.  We also examine the instability
found in Ref.~\cite{Bojowald:2007ra} and show numerically that the
analytic conditions for stability found for a specific lattice
refinement model, are indeed valid. We then look at how these
conditions change with different lattice refinement models. Finally,
we comment of the fact that the discrete nature of the underlying
lattice introduces a {\it twist} into otherwise straight Gaussian
wave-packets and explain its origin.

\section{Elements of Loop Quantum Cosmology}\label{sec:background}
Loop quantum cosmology is not formulated in terms of metrics and
coordinates, rather SU(2) holonomies of the connection, $\hat{h}_k$,
and triads, $\hat{p}$, are used. In this set up the gravitational part of
the Hamiltonian constraint\footnote {The reader should note that the
factor ordering we are using is not the {\sl conventional} one. We
make this choice because the Hamiltonian constraint for a lattice
refinement model of the form $\tilde\mu=\mu_0\mu^{-A}$ with general
$A$, turns out to be of the same form as that for the fixed lattice
case. This simplifies the demonstration of our numerical method to
solve the quantum evolution equation of generic lattice-refined
models. Certainly, the validity of our method is independent of the
choice for the factor ordering.}, assuming flat homogeneous and
isotropic models, reads~\cite{Ashtekar:2006wn}
\be 
\hat{\cal C}_{\rm grav} |\Psi\rangle = \frac{2i}{\kappa^2 \hbar
  \gamma^3 \tilde{\mu}^3} {\rm tr} \sum_{ijk} \epsilon^{\rm ijk}
\left( \hat{h}_k \left[ \hat{h}_k^{-1}, \hat{V} \right]\hat{h}_ i
\hat{h}_j \hat{h}_i^{-1} \hat{h}_j^{-1} \right) |\Psi \rangle~,
\label{ham-constr}
\ee
where $\kappa = 8\pi G$ and $\hat{V} = \widehat{\left| p \right|^{3/2}
}$ is the volume operator.  We use the irreducible representation,
$J=1/2$, so that the Hamiltonian constraint is not plagued of
ill-behaving spurious solutions~\cite{Vandersloot:2005kh}.
The action of the volume operator $\hat V$
on the basis states $|\mu \rangle$ is given by
\be
\hat{V}|\mu\rangle =\widehat{|p|^{3/2}}|\mu\rangle = 
\left( \frac{\kappa \gamma \hbar |\mu |}{6} \right)^{3/2} |
\mu \rangle~.
\ee
The basis states $|\mu \rangle$ are eigenstates of the triad, and
eigenstates of the volume operator, with eigenvalues $\mu$. They
satisfy the orthonormality relation
\be
\langle\mu_1|\mu_2\rangle=\delta_{\mu_1,\mu_2}~.
\ee 
The holonomy $h_i$ (more precisely $h_i^{(\tilde\mu)}$) is along the
edge parallel to the $i^{(\rm th)}$ basis vector, whose length is set
by $\tilde\mu$.  The parameter $\gamma$ is the Barbero-Immirzi
parameter, a constant ambiguity parameter that can be fixed by
considering black-hole entropy
calculations~\cite{Ashtekar:2000eq,Ashtekar:2003zx}.  Classically,
$p=a^2$, where $a$ is the usual cosmological scale factor.

A general state $|\Psi\rangle$ in the kinematical Hilbert space can
be expanded as
\be
|\Psi\rangle = \sum_\mu \Psi_\mu |\mu\rangle~,
\ee
with the requirement
\be
\sum_\mu \bar\Psi_\mu \Psi_\mu<\infty~,
\ee
so that the state has a finite kinematical norm.

The action of the holonomies of the Ashtekar connection on the
basis states reads~\cite{Ashtekar:2006wn}
\be
\hat{h}_i|\mu\rangle= (\widehat{\rm cs}{\mathbb 1} -i \sigma _i
\widehat{\rm sn})|\mu\rangle~,
\ee
where $\mathbb 1$ is the identity $2\times 2$ matrix, $\sigma _i$
are the Pauli spin matrices and $\widehat{\rm cs}$, $\widehat{\rm
  sn}$ are given by
\beq
\widehat{\rm cs} |\mu\rangle &\equiv& \frac{1}{2}\left( \widehat{
   e^{i\tilde{\mu}\bar{c}/2} } + \widehat{ e^{-i\tilde{\mu}\bar{c}/2}
 } \right)|\mu\rangle = \frac{1}{2} \left( |\mu+\tilde{\mu}\rangle +
 |\mu-\tilde{\mu}\rangle \right) \nonumber \\ 
\widehat{\rm sn} |\mu\rangle &\equiv& \frac{1}{2}\left( \widehat{
  e^{i\tilde{\mu}\bar{c}/2} } - \widehat{ e^{-i\tilde{\mu}\bar{c}/2} }
\right)|\mu\rangle = \frac{1}{2i} \left( |\mu+\tilde{\mu}\rangle -
|\mu-\tilde{\mu}\rangle \right)~, 
\label{eq:hol}
\eeq
with $\tilde\mu$ a real number.

The generalised isotropic connection $|\bar{c}\rangle$ is
defined such that, 
\be
\langle \bar{c} | \mu\rangle = e^{i\mu\bar{c}/2}~.
\ee
Using Eq.~(\ref{eq:hol}), the gravitational part of the Hamiltonian
constraint, Eq.~(\ref{ham-constr}), can be written in the form of a
difference equation
\be
\label{eq:grav_diff} 
\hat{\cal C}_{\rm grav} |\Psi \rangle = \frac{3}{2\kappa^2\hbar\gamma^3
\tilde\mu^2} \sum_\mu S(\mu) \left[
  \Psi_{\mu+4\tilde{\mu}} - 2\Psi_\mu +\Psi_{\mu-4\tilde{\mu}} \right]
|\mu \rangle~,
\ee
where
\beq
 S(\mu) &\equiv& \frac{2}{\gamma\kappa\hbar\tilde\mu} |
V_{\mu + \tilde{\mu}} - V_{\mu-\tilde{\mu}}|~. \nonumber \\
&=&\frac{1}{3\tilde{\mu}}\sqrt{\frac{\kappa \gamma \hbar}{6} } \left| 
\left| \mu+\tilde{\mu}
\right|^{3/2} - \left| \mu - \tilde{\mu} \right|^{3/2}\right|~.
\eeq
From now on we set, for simplicity, $1/(2\sqrt{6\kappa^3\hbar\gamma^5})=1$,
so that the gravitational part of the Hamiltonian, Eq.~(\ref{eq:grav_diff}) is,
\be\label{eq:grav_diff2}
\hat{\cal C}_{\rm grav} |\Psi \rangle = \sum_\mu \frac{1}{\tilde{\mu}^3}\left| 
\left| \mu+\tilde{\mu}
\right|^{3/2} - \left| \mu - \tilde{\mu} \right|^{3/2}\right| \left[
  \Psi_{\mu+4\tilde{\mu}} - 2\Psi_\mu +\Psi_{\mu-4\tilde{\mu}} \right]
|\mu \rangle~,
\ee
and the full Hamiltonian constraint reads
\be
\label{eq:ham}
\left( \hat{\cal C}_{\rm grav} + \hat{\cal H}_\phi \right)|\Psi\rangle = 0~,
\ee 
where $\hat{\cal H}_\phi$ is the matter Hamiltonian, assumed to
operate diagonally on the basis states, i.e., $\hat{\cal H}_\phi
|\mu\rangle = {\cal H}_\phi |\mu\rangle$. Using
Eq.~(\ref{eq:grav_diff2}), the full Hamiltonian constraint,
Eq.~(\ref{eq:ham}), reads 
\be\label{eq:vary1}
\frac{1}{\tilde{\mu}^3}\left| \left| \mu+\tilde{\mu}
\right|^{3/2} - \left| \mu - \tilde{\mu} \right|^{3/2}\right|
 \left[ \Psi_{\mu+4\tilde{\mu}} - 2\Psi_\mu +\Psi_{\mu-4\tilde{\mu}}
 \right] = -{\cal H}_\phi~,
\ee
Until recently, the magnitude of the shift
operator, $\tilde{\mu}$, was taken to be a constant. This gave the
resulting difference equation a fixed step-size making its analysis
much simpler. However, in the full theory one expects that
$\tilde{\mu}$ will be a decreasing function of $\mu$ and hence the
step-size in the difference equation, Eq.~(\ref{eq:vary1}), will vary.
In essence, the difference equation is defined on a {\it refining
  lattice}. It has been shown that modelling of this lattice
refinement is crucial for the stability of the classical cosmological
wave-function~\cite{Bojowald:2007ra,Nelson:2007wj,Nelson:2007um}.

In the case of an one-dimensional system, such as the one under
consideration here, the problem can be mapped onto a fixed lattice
simply by a change of basis. This depends somewhat on the precise form
of lattice refinement, however all that is required for it to be
possible is that the integral $\int \tilde{\mu}\left(\mu\right) d\mu$
exists. To be explicit, consider a lattice refinement model of the
form $$\tilde{\mu} = \mu_0 \mu^{-A}~,$$ where $\mu_0$ is some
constant~\cite{Nelson:2007um}.  If we then make the change of
variables $$\mu \rightarrow \nu = k\frac{\mu^{1-A}}{\mu_0(1-A)}~,$$
where $k$ is a constant, equal to the magnitude of the shift operator
associated with these new coordinates, Eq.~(\ref{eq:vary1}) becomes
\be \frac{1}{k^3} S(\nu) \left[ \Psi_{\nu+4k} - 2\Psi_\nu
  +\Psi_{\nu-4k} \right] = -{\cal H}_\phi~, \label{eq:const1} \ee
where \be |\Psi\rangle = \sum_\nu \psi_\nu |\nu\rangle~, \ee and
\be S(\nu) = \Bigl| \left| \left( \nu + k\right) \alpha \right|^{3/
  2/(1-A)} - \left| \left( \nu - k\right) \alpha \right|^{3/2/(1-A)}
\Bigr|~, \ee with $\alpha$ defined
as $$\alpha\equiv\frac{\mu_0(1-A)}{k}~.$$ The fact that the difference
equations, Eq.~(\ref{eq:vary1}) and Eq.~(\ref{eq:const1}), are of the
same form is due to the choice of factor ordering. This allows us to
directly compare numerical solutions of the two systems, since they
will have the same large-scale limit. This is indeed the reason for
our choice of factor ordering in Eq.~(\ref{ham-constr}).

Note that usually one wishes to use a self-adjoint Hamiltonian,
however for simplicity (and to allow for a more direct comparison with
the case of a black hole interior), we use the form given in
Eq.~(\ref{eq:grav_diff}), which is not self-adjoint. Certainly, the
numerical method described below is also valid for the self-adjoint
case, as well as for different factor orderings.

\section{Numerical evolution of the difference equation}
\subsection{One-dimensional case}
\begin{figure}
  \begin{center}
\includegraphics[scale=0.85]{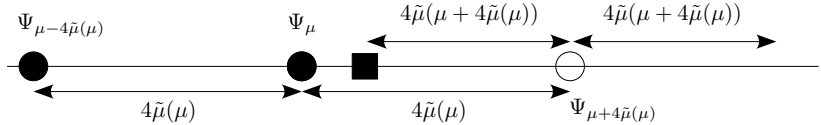}
 \caption{\label{fig1} This figure demonstrates the problem of
   evolving a difference equation that has been defined on a varying
   lattice. Given the value of $\Psi$ defined on two adjacent lattice
   points (solid circles) one can use the difference equation to
   evaluate $\Psi$ on the subsequent lattice site (open
   circle). However when one attempts to evaluate $\Psi$ at the next
   lattice point, one finds that one does not have the necessary data
   point (square).}
  \end{center}
\end{figure}

In the previous section we have seen how to transform the Hamiltonian
constraint from an one-dimensional difference equation defined on a
varying lattice, Eq.~(\ref{eq:vary1}), to one on a constant lattice,
Eq.~(\ref{eq:const1}). Thus, given the initial values of $\Psi$ on two
adjacent lattice points, one can iterate Eq.~(\ref{eq:const1}) to
calculate $\Psi$ on all lattice points. However, this mapping of the
problem onto a constant lattice is not, in general, possible in two
dimensions; one needs to develop a system that allows the use of the
difference equation defined on the varying lattice,
Eq.~(\ref{eq:vary1}).  It is clear that, given two values of $\Psi$
defined on two adjacent lattice points, one can no longer iterate the
difference equation to arbitrary $\mu$~\cite{Sabharwal:2007xy} ({\sl
  see} Fig.~\ref{fig1} which illustrates this problem).
Assuming that the wave-function is {\it
  pre-classical}~\cite{Bojowald:2002ny}, i.e., that it varies slowly on
scales smaller than the discreteness scale, one can use a Taylor
expansion about previously calculated lattice sites to approximate the
data necessary for the next iteration. This is equivalent to the local
interpolation method used in Ref.~\cite{Sabharwal:2007xy}. The main
advantage however of the Taylor expansion method we propose here is
that it allows one to estimate the order of the approximation and, if
necessary, increase the accuracy.

Being more explicit, consider the scheme depicted in Fig.~\ref{fig1},
and set
$$\Psi_{\mu-4\tilde{\mu}\left(\mu\right)}=\Psi_1\ ,
\ \Psi_{\mu}=\Psi_2\ , \ \Psi_{ \mu+4\tilde{\mu}\left(\mu\right)}
=\Psi_3~.$$ Given the value of $\Psi_1$ and $\Psi_2$, one can use
Eq.~(\ref{eq:vary1}) to evaluate $\Psi_3$. We then move to the next
lattice point, so that $\bar{\Psi}_2 = \Psi_3$, where the over-line
indicates this is the `new' value. To calculate $\bar{\Psi}_1$ we make
a Taylor expansion about $\Psi_2$ to get
\beq
 \bar{\Psi}_1 &=& \Psi_2 + \frac{1}{2}\left[ \Psi_3 -
   2\Psi_2+\Psi_1\right]\left[ 1-\Delta(\mu) \right] \nonumber 
\\ &&+ {\cal O} \left[ [\tilde{\mu}\left(\mu\right)]^2 \left[ 1 -
         \Delta(\mu) \right] \frac{{\rm d}^2 \Psi}{{\rm
           d}\mu^2}\Big|_\mu \right] + {\cal O} \left[
       [\tilde{\mu}\left(\mu\right)]^2\frac{{\rm d}^3 \Psi}{{\rm
           d}\mu^3} \Big|_\mu \right]~,
\eeq
where
\be
\Delta(\mu) =
\frac{\tilde{\mu}(\mu_1)}{\tilde{\mu}(\mu)}~,\ \ \mbox{with}\ \ \mu_1=
\mu+4\tilde\mu(\mu)~.
\ee 
Here, we have only used the first-order terms in the expansion to
calculate the required data point, with the second-order terms being
used to keep track of the accuracy of the approximation. This system
readily extends to higher order. It should be noted that if higher
order terms are required, then Eq.~(\ref{eq:vary1}) must be used to
evaluate terms like $\Psi_{\mu+8\tilde{\mu}(\mu)}$,
$\Psi_{\mu+12\tilde{\mu}(\mu)}$, etc., so that the higher derivatives
can be calculated.

\begin{figure}
  \begin{center}
\includegraphics[scale=0.85]{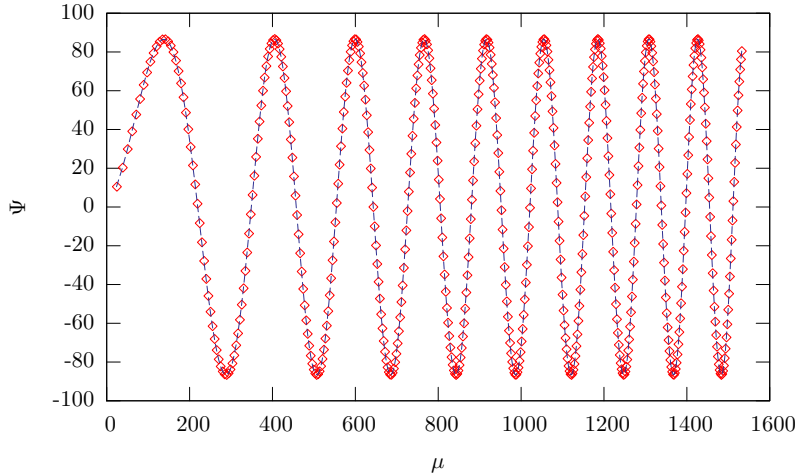}
 \caption{\label{fig2} The evolution of a wave-function using the
   Taylor expansion scheme. The circles are the approximated values
   calculated on the varying lattice, whilst the line is the exact
   wave-function. We use $\mu_0 = 1.0$, $A=-0.5$, ${\cal H}_\phi = 1.0\times
   10^{-4}$, $\mu_{\rm initial}=1.0$, and initial conditions $\Psi_1 =
   0.0$, $\Psi_2 = 0.5$.}
  \end{center}
\end{figure}
\begin{figure}
  \begin{center}
\includegraphics[scale=0.85]{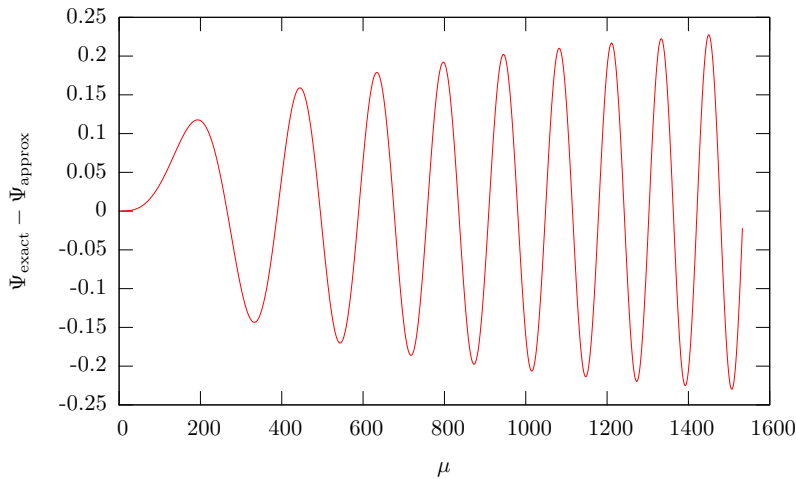}
 \caption{\label{fig3} The difference between the exact approximate
   wave-functions is less than one percent. This error is almost
   entirely due to the slight difference in the initial conditions
   between the two grids, i.e., the difference between the initial
   separation of $\Psi_1$ and $\Psi_2$. We use $\mu_0 = 1.0$,
   $A=-0.5$, ${\cal H}_\phi = 1.0\times 10^{-4}$, $\mu_{\rm
     initial}=1.0$, and initial conditions $\Psi_1 = 0.0$, $\Psi_2 =
   0.5$.}
  \end{center}
\end{figure}
\begin{figure}
  \begin{center}
\includegraphics[scale=0.85]{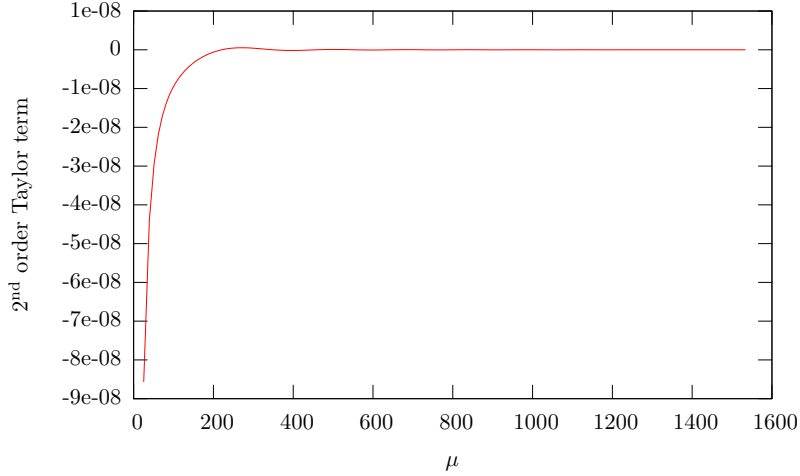}
 \caption{\label{fig4} The value of the next order term in the Taylor
   expansion. We use $\mu_0 = 1.0$, $A=-0.5$, ${\cal H}_\phi =
   1.0\times 10^{-4}$, $\mu_{\rm initial}=1.0$, and initial conditions
   $\Psi_1 = 0.0$, $\Psi_2 = 0.5$.}
  \end{center}
\end{figure}

As an example, we evaluate the wave-function using both, the exact
case, Eq.~(\ref{eq:const1}), and Taylor expansion method described
above. A small, constant matter Hamiltonian is used (${\cal H}_\phi =
1.0\times 10^{-4}$), to give the wave-functions some fine detail,
however it should be remembered that there is now the possibility that
gravitational back reaction may become important.
The resulting wave-functions, obtained with both approaches, are given
in Fig.~\ref{fig2}; their difference is plotted in
Fig.~\ref{fig3}. The next order term in the Taylor expansion is shown
in Fig.~\ref{fig4}. It is clear that, even at linear order, the Taylor
expansion method is extremely accurate. The difference between the two
wave-functions is almost entirely due to the fact that the separation
between the two initial lattice points is not exactly the same, as a
result of the lattice refinement of the $\tilde{\mu}\left(\mu\right)$
scheme. This alters the initial conditions slightly, nevertheless the
discrepancy is still less than one percent.

It should be noted that if the lattice is not refining fast enough to
ensure that the wave-function remains {\sl pre-classical}, the
interpolation method would begin to
fail~\cite{Nelson:2007um}. However, this would correspond to an
unstable wave-function which would not have a classical large-scale
limit.

\subsection{Two-dimensional case}

The cosmological quantisation procedure used in
Section \ref{sec:background} can be adapted to the anisotropic
geometry of a black hole interior. The resulting two-dimensional
Hamiltonian constraint is again a difference equation defined on a
varying lattice~\cite{Bojowald:2007ra},
\beq
\label{eq:vary2D}
&&C_{+}\left(\mu,\tau\right) \left[
  \Psi_{\mu+2\delta_\mu,\tau+2\delta_\tau} - \Psi_{\mu-2\delta_\mu,
    \tau+2\delta_\tau} \right] \nonumber \\ 
&&+ C_0\left(\mu,\tau\right) \left[
  \left(\mu+2\delta_\mu\right)\Psi_{\mu+4\delta_\mu, \tau} - 2\left( 1
  + 2 \gamma^2 \delta_\mu^2\right) \mu\Psi_{\mu,\tau}
  +\left(\mu-2\delta_\mu\right)\Psi_{\mu-4\delta_\mu,\tau} \right]
\nonumber \\ 
&&+C_{-} \left( \mu,\tau\right) \left[
  \Psi_{\mu-2\delta_\mu,\tau-2\delta_\tau} -
  \Psi_{\mu+2\delta_\mu,\tau-2\delta_\tau}\right] = \frac{\delta_\tau
  \delta_\mu^2}{\delta^3} {\cal H}_\phi\Psi_{\mu,\tau}~, 
\eeq 
with
\beq 
C_{\pm} &\equiv& 2\delta_\mu \left( \sqrt{\left| \tau \pm 2
  \delta_\tau \right| } + \sqrt{\left| \tau\right| } \right)~, \\ 
C_0 &\equiv& \sqrt{\left| \tau + \delta_\tau\right|} - \sqrt{ \left| \tau -
  \delta_\tau\right| }~, 
\eeq 
where we have defined $\delta_\mu$ and $\delta_\tau$ as the step-sizes
along the $\mu$ and $\tau$ directions, respectively.  The parameter
$\delta$, with $0<\delta<1$, gives the fraction of a lattice edge that
the underlying graph changing Hamiltonian
uses~\cite{Bojowald:2007ra}. For clarity, the $\mu$ and $\tau$
dependence in $\delta_\mu\left(\mu,\tau\right)$ and
$\delta_\tau\left(\mu,\tau\right)$ have been suppressed.  The lattice
spans the $\left(\mu,\tau\right)$-plane and since $\mu$ and $\tau$ are
the coordinate lengths along the polar and radial coordinates,
respectively, the $\left(\mu,\tau\right)$-plane corresponds to the
$\left( \Theta, r\right)$-plane of the black-hole interior.

\begin{figure}
 \begin{center}
   \includegraphics{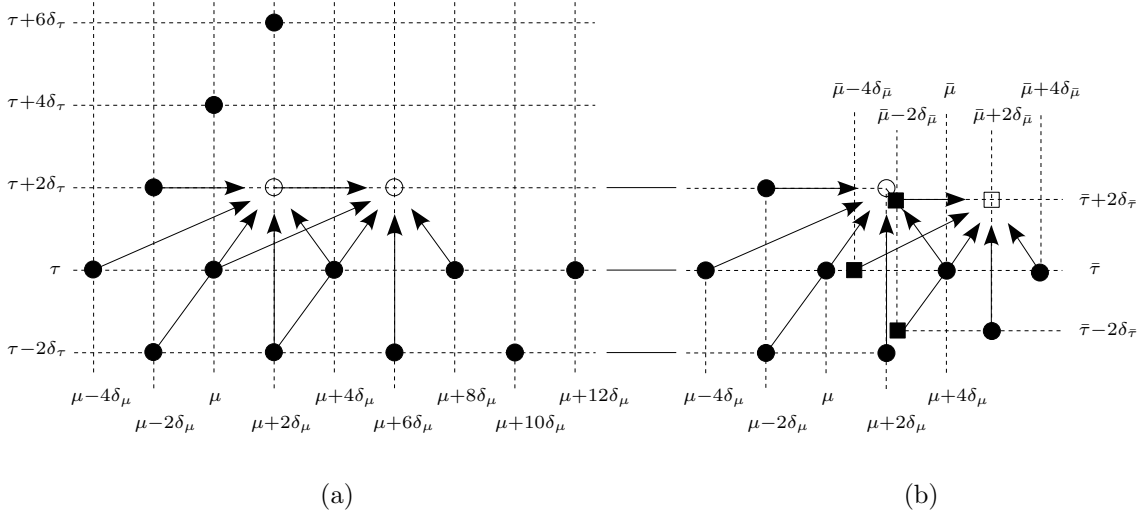}
   \caption{\label{fig5} (a) For the fixed lattice case the
     two-dimensional wave-function can be calculated, given suitable
     initial conditions (solid circles).  (b) In the case of a
     refining lattice, the data needed to calculate the value of the
     wave-function at a particular lattice site (open square) is not
     given by previous iterations (solid squares). This is due to the
     fact that $\delta_\mu(\mu_i,\tau_i)-\delta_\mu(\mu_{i+1},
     \tau_i)\neq 0$, where $\mu_{i+1}=\mu_i+4\delta_\mu
     (\mu_i,\tau_i)$. To improve the clarity of the diagram, the
     explicit $\mu$ and $\tau$ dependence of the step-sizes is
     suppressed, and the notation
     $\delta_{\bar{\mu}}\equiv\delta_{\mu}(\mu_{i+1},\tau_i)$, and
     $\delta_{\bar{\tau}}\equiv\delta_{\tau}(\mu_{i+1},\tau_i) $ is
     used.}
 \end{center}
\end{figure}

\begin{figure}
 \begin{center}
  \includegraphics[width=8.7cm]{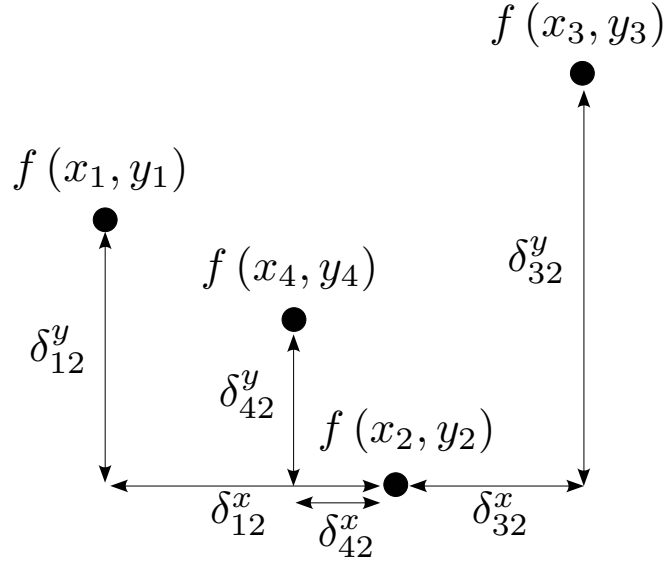}
  \caption{\label{fig6} Given the value of a function evaluated at
    three positions, $\left(x_1,y_1\right)$, $\left(x_2,y_2\right)$ and
    $\left(x_3,y_3\right)$, a Taylor expansion can be used to
    approximate the value at a fourth point, $\left(x_4,y_4\right)$.}
 \end{center}
\end{figure}

Note that we have again assumed that the matter Hamiltonian acts
diagonally on the basis states of the wave-function, namely 
\be
\hat{\cal H}_\phi |\Psi \rangle \equiv \hat{\cal H}_\phi
\sum_{\mu,\tau} \Psi_{\mu,\tau} |\mu,\tau\rangle = \sum_{\mu,\tau}
    {\cal H}_\phi \Psi_{\mu,\tau} |\mu,\tau \rangle~.  
\ee
If $\delta_\mu$ and $\delta_\tau$ were constant, then
Eq.~(\ref{eq:vary2D}) could be used to iteratively calculate the value
of $\Psi$ at each successive lattice point, given suitable initial
conditions ({\sl see}, Fig.~\ref{fig5}a). If instead the lattice is
refining, i.e.,  if $\delta_\mu$ and $\delta_\tau$ are decreasing
functions of $\mu$ and $\tau$, respectively, then we have the same
problem as in the one-dimensional case ({\sl see}, Fig.~\ref{fig5}b).
As in the one-dimensional case, one can use Taylor expansions to
calculate the necessary data points.  In general, given a function
evaluated at three (non-co-linear) coordinates, the Taylor
approximation to the value at a fourth position is given by
\beq
 f\left( x_4,y_4\right) &=& f\left(x_2, y_2\right) + \delta^x_{42}
 \frac{\partial f}{\partial x} \Big|_{x_2, y_2} + \delta^y_{42}
 \frac{\partial f}{\partial y} \Big|_{x_2, y_2} \nonumber\\ 
&& + {\cal O}\left( \left(\delta^x_{42}\right)^2 \frac{\partial^2
   f}{\partial x^2}\Big|_{x_2,y_2}\right)+{\cal O}\left(
 \left(\delta^y_{42}\right)^2 \frac{\partial^2 f}{\partial
   y^2}\Big|_{x_2,y_2}\right) ~,\label{eq:taylor} 
\eeq 
where the Taylor expansion is taken about the position
$\left(x_2,y_2\right)$ and we have defined $\delta_{ij}^x \equiv x_i-x_j$ and
$\delta_{ij}^y \equiv y_i-y_j$ ({\sl see}, Fig.~\ref{fig6}).
To approximate the differentials in Eq.~(\ref{eq:taylor}), use points
$\left(x_1,y_1\right)$ and $\left( x_3,y_3\right)$:
\beq
f\left(x_1,y_1\right) &=& f\left(x_2,y_2\right) + \delta^x_{12} \frac{
  \partial f}{\partial x}\Big|_{ x_2,y_2} + \delta^y_{12}
\frac{\partial f}{\partial y}\Big|_{x_2,y_2} + \cdots~,\\
f\left(x_3,y_3\right) &=& f\left(x_2,y_2\right) + \delta^x_{32} \frac{
  \partial f}{\partial x}\Big|_{ x_2,y_2} + \delta^y_{32}
\frac{\partial f}{\partial y}\Big|_{x_2,y_2} + \cdots~,
\eeq
where the dots indicate higher order terms. Solving for $\partial_x f$
and $\partial_y f$, gives
\beq
 \frac{\partial f}{\partial x}\Big|_{x_2,y_2} &=&
 \frac{\delta^y_{12}}{\Delta} \left[ f\left( x_3,y_3\right) -
   f\left(x_2,y_2\right) \right] - \frac{ \delta^y_{32}}{\Delta}
 \left[ f\left( x_1,y_1\right)- f\left(x_2, y_2\right)
   \right]+\cdots~, \nonumber\\
\frac{\partial f}{\partial y}\Big|_{x_2,y_2} &=&
\frac{\delta^x_{32}}{\Delta} \left[ f\left( x_1,y_1\right) -
  f\left(x_2,y_2\right) \right] - \frac{ \delta^y_{12}}{\Delta} \left[
  f\left( x_3,y_3\right)- f\left(x_2, y_2\right) \right]+\cdots~,
\nonumber 
\eeq
where $\Delta\equiv \delta^x_{32}\delta^y_{12} -
\delta^y_{32}\delta^x_{12}$.

As in the one-dimensional case, higher-order terms in the Taylor
expansion can be used to improve the accuracy of the system. Here we
calculate the second-order expansion to demonstrate that linear
interpolation is sufficient for many interesting cases. Given five
points $\Psi_1$, $\Psi_2$, $\Psi_3$, $\Psi_4$ and $\Psi_5$, where
$\Psi_{\rm n} = \Psi\left(\mu_{\rm n},\tau_{\rm n}\right)$, the
second-order Taylor approximation for the sixth point, $\Psi_6$, reads
\be
\label{eq:2nd_order}
 \Psi_6 = \Psi_1 +
 \delta^\mu_{61}\frac{\partial\Psi}{\partial\mu}\Big|_1 +
 \delta^\tau_{61}\frac{\partial\Psi}{\partial\tau}\Big|_1 +
 \frac{\left(\delta^\mu_{61}\right)^2}{2}\frac{\partial^2\Psi}
{\partial\mu^2}\Big|_1 +
 \frac{\left(\delta^\tau_{61}\right)^2}{2}\frac{\partial^2\Psi}
{\partial\tau^2}\Big|_1 +\cdots~, 
\ee 
where as before $\delta^\mu_{ij} \equiv \mu_i - \mu_j$ and
$\delta^\tau_{ij}\equiv\tau_i - \tau_j$; the
derivatives are all evaluated at the point $\left(\mu_1,
\tau_1\right)$. We can calculate the derivatives to the necessary
order by solving the following system of simultaneous equations:
\beq
 \Psi_2 = \Psi_1 +
 \delta^\mu_{21}\frac{\partial\Psi}{\partial\mu}\Big|_1 +
 \delta^\tau_{21}\frac{\partial\Psi}{\partial\tau}\Big|_1 +
 \frac{\left(\delta^\mu_{21}\right)^2}{2}\frac{\partial^2\Psi}
{\partial\mu^2}\Big|_1 +
 \frac{\left(\delta^\tau_{21}\right)^2}{2}\frac{\partial^2\Psi}
{\partial\tau^2}\Big|_1~, \nonumber \\ 
\Psi_3 = \Psi_1 +
 \delta^\mu_{31}\frac{\partial\Psi}{\partial\mu}\Big|_1 +
 \delta^\tau_{31}\frac{\partial\Psi}{\partial\tau}\Big|_1 +
 \frac{\left(\delta^\mu_{31}\right)^2}{2}\frac{\partial^2\Psi}
{\partial\mu^2}\Big|_1 +
 \frac{\left(\delta^\tau_{31}\right)^2}{2}\frac{\partial^2\Psi}
{\partial\tau^2}\Big|_1~, \nonumber \\ 
\Psi_4 = \Psi_1 +
 \delta^\mu_{41}\frac{\partial\Psi}{\partial\mu}\Big|_1 +
 \delta^\tau_{41}\frac{\partial\Psi}{\partial\tau}\Big|_1 +
 \frac{\left(\delta^\mu_{41}\right)^2}{2}\frac{\partial^2\Psi}
{\partial\mu^2}\Big|_1
 +
 \frac{\left(\delta^\tau_{41}\right)^2}{2}\frac{\partial^2\Psi}
{\partial\tau^2}\Big|_1~, \nonumber \\ 
\Psi_5 = \Psi_1 +
 \delta^\mu_{51}\frac{\partial\Psi}{\partial\mu}\Big|_1 +
 \delta^\tau_{51}\frac{\partial\Psi}{\partial\tau}\Big|_1 +
 \frac{\left(\delta^\mu_{51}\right)^2}{2}\frac{\partial^2\Psi}{
\partial\mu^2}\Big|_1
 +
 \frac{\left(\delta^\tau_{51}\right)^2}{2}\frac{\partial^2\Psi}
{\partial\tau^2}\Big|_1~.
 \eeq 
One gets
\be 
\frac{\partial\Psi}{\partial \mu}\Big|_1 =
 \frac{\Delta_1}{\Delta_0}~, \ \ \ \ \frac{\partial\Psi}{\partial
   \tau}\Big|_1 = - \frac{\Delta_2}{\Delta_0}~,
 \ \ \ \ \frac{\partial^2\Psi}{\partial\mu^2}\Big|_1 =
 \frac{\Delta_3}{\Delta_0}~,
 \ \ \ \ \frac{\partial^2\Psi}{\partial\tau^2}\Big|_1 =
 - \frac{\Delta_4}{\Delta_0}~, 
\ee 
where $\Delta_{\rm i}$ is the determinant of, 
\be
\left( \begin{array}{ccccc}
   \Psi_1 - \Psi_2& \delta^\mu_{21} & \delta^\tau_{21} 
& \left(\delta^\mu_{21}\right)^2/2 &
  \left(\delta^\tau_{21}\right)^2/2  \\
   \Psi_1 - \Psi_3 & \delta^\mu_{31} & \delta^\tau_{31} 
& \left(\delta^\mu_{31}\right)^2/2 &
 \left(\delta^\tau_{31}\right)^2/2\\
   \Psi_1 - \Psi_4 & \delta^\mu_{41} & \delta^\tau_{41} 
& \left(\delta^\mu_{41}\right)^2/2 &
 \left(\delta^\tau_{41}\right)^2/2\\
   \Psi_1 - \Psi_5 & \delta^\mu_{51} & \delta^\tau_{51} 
& \left(\delta^\mu_{51}\right)^2/2 &
 \left(\delta^\tau_{51}\right)^2/2
   \end{array} \right)~,
\ee
with the $(i+1)^{\rm th}$ column removed. Substituting these
approximations for the differentials into Eq.~(\ref{eq:2nd_order}),
one obtains the second-order Taylor approximation to the point
$\Psi\left(\mu_6,\tau_6\right)$, as required.

\begin{figure}
 \begin{center}
  \includegraphics[scale=0.75]{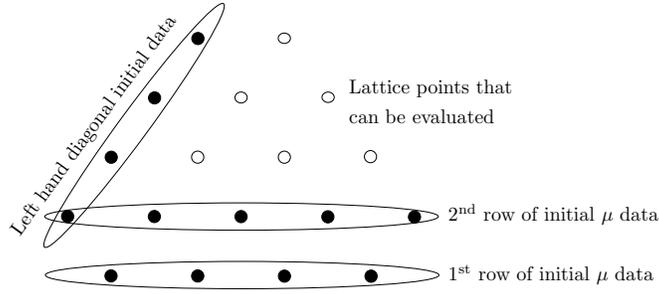}
  \caption{\label{fig14} The initial data (solid circles) is given on
    two adjacent $\mu$ rows and the lattice points along the left-hand
    (small $\tau$) diagonal. This allows us to evaluate the
    wave-function on lattice points (open circles) within a triangular
    region of base $\mu_{\rm max} - \mu_{\rm initial}$ and height
    $\tau_{\rm max} - \tau_{\rm initial}$.}
 \end{center}
\end{figure}

\begin{figure}
 \begin{center}
  \includegraphics{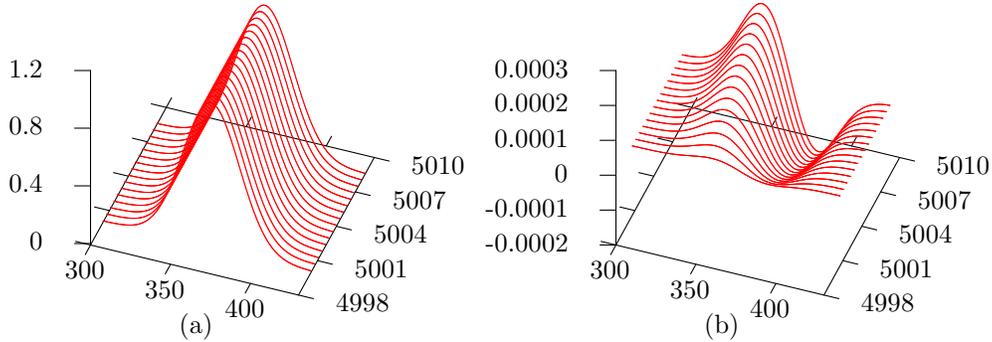}
  \caption{\label{fig7} The wave-function is calculated by iterating
    the difference equation using first-order Taylor expansion to
    evaluate the necessary data points at each site. The initial
    conditions are a Gaussian on $\tau_{\rm initial} = 5000$, centred
    between $\mu_{\rm initial}= 100$ and $\mu_{\rm max} = 735$, of
    width $\sigma = 25$. The lattice refinement model used is
    $\delta_\mu\left(\mu,\tau\right)=\mu^{-1/2}$,
    $\delta_\tau\left(\mu,\tau\right) = \tau^{-1/2}$.  Only the
    relevant section of the wave-function is plotted and to improve
    clarity, only every $20^{\rm th}$-$\tau$ lattice point is
    shown. In (a) the propagation of the the wave-function is shown,
    whilst in (b) the value of the second-order terms is
    plotted. The second-order corrections are typically of
    the order of $10^{-2}~\%$ over this range.}
 \end{center}
\end{figure}

This system has been implemented for the vacuum case (${\cal H}_\phi
=0$) for an initially Gaussian wave-packet along the $\mu$-direction.
As shown in Fig.~\ref{fig14}, the initial data consist of two
$\mu$-rows of data, adjacent in the $\tau$-direction and the data
along the left-hand diagonal.
One can also see from Fig.~\ref{fig14} that only data points within a
similar diagonal on the right-hand side can be calculated. This
results in the $\left(\mu,\tau\right)$ being restricted to a triangle
of base $\mu_{\rm max} - \mu_{\rm initial}$ and height $\tau_{\rm max}
- \tau_{\rm initial}$. Figure \ref{fig7}a shows a typical output using
the first-order approximation evaluated for the lattice refinement
model $\delta_\mu\left(\mu,\tau\right)=\mu^{-1/2}$,
$\delta_\tau\left(\mu,\tau\right) = \tau^{-1/2}$, whilst
Fig.~\ref{fig7}b gives the second-order correction to this. It is
clear that, at least for slowly varying wave-functions, the linear
approximation is extremely accurate (higher-order corrections being
$\approx 10^{-2}~\%$).

\section{Stability of the Schwarszchild interior}
In Ref.~\cite{Bojowald:2007ra} a von Neumann stability analysis of the
difference equation, Eq.~(\ref{eq:vary2D}), was performed for two
lattice refinement models, and it was shown that in certain
circumstances the system is only conditionally stable. In particular,
it was found that for $\delta_\mu\left(\mu,\tau\right) =
\mu_0\mu^{-1}$ and $\delta_\tau\left(\mu,\tau\right) = \tau_0
\tau^{-1}$, the system is unstable for $\mu>2\tau$. 

\begin{figure}
 \begin{center}
  \includegraphics{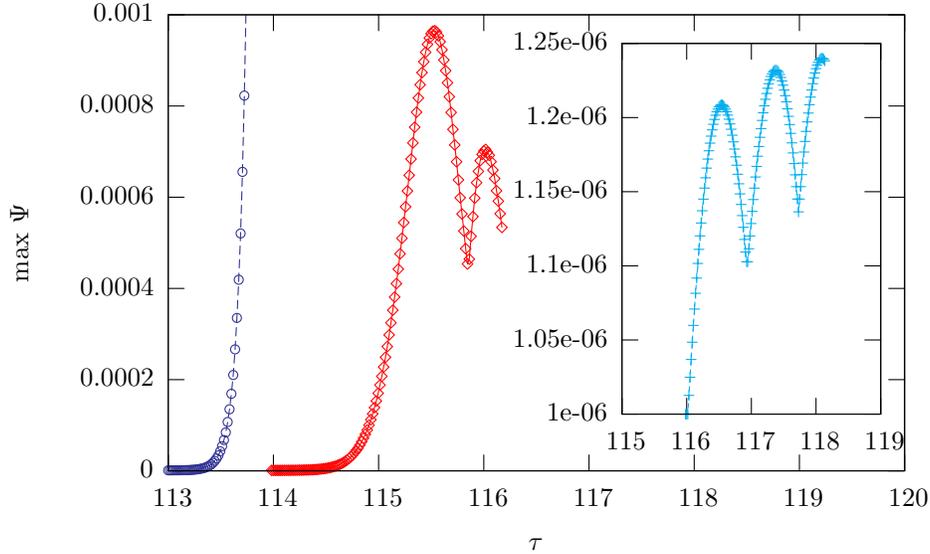}
  \caption{\label{fig9} A perturbation of $\Psi = 10^{-6}$ was put on
    an otherwise empty initial $\mu$-row at $\mu = 230$, for the
    lattice refinement model $\delta_\mu\left(\mu,\tau\right) =
    \mu^{-1}$, $\delta_\tau\left(\mu,\tau\right)=\tau^{-1}$. The
    maximum amplitude of the resulting wave-function is plotted as a
    function of $\tau$, for the initial perturbation set at $\tau =
    113$, $\tau = 114$ and $\tau = 116$. Analytically, the region of
    stability is given for $\tau > \mu/2$, i.e., the amplitude of the
    perturbation grows exponentially for $\tau < 115$ and oscillates
    for $\tau >115$. Here we confirm this numerically. Note the
    different scale on the graph starting within the stable region. }
 \end{center}
\end{figure}

We investigate this instability numerically and show that the
stability condition is indeed correct. We do this by using the scheme
described above to evaluate the wave-function, given two initial
(consecutive) $\mu$ rows, that are empty apart from a small
($10^{-6}$) perturbation at a particular value of $\mu$. The
perturbation needs to be small to ensure we remain in the {\sl
  pre-classical} regime. The system is then evaluated according to the
the difference equation, Eq.~(\ref{eq:vary2D}), across different
ranges of $\tau$. Figure~(\ref{fig9}) shows a typical example of how
the amplitude of the perturbation varies with $\tau$.

\begin{figure}
 \begin{center}
  \includegraphics{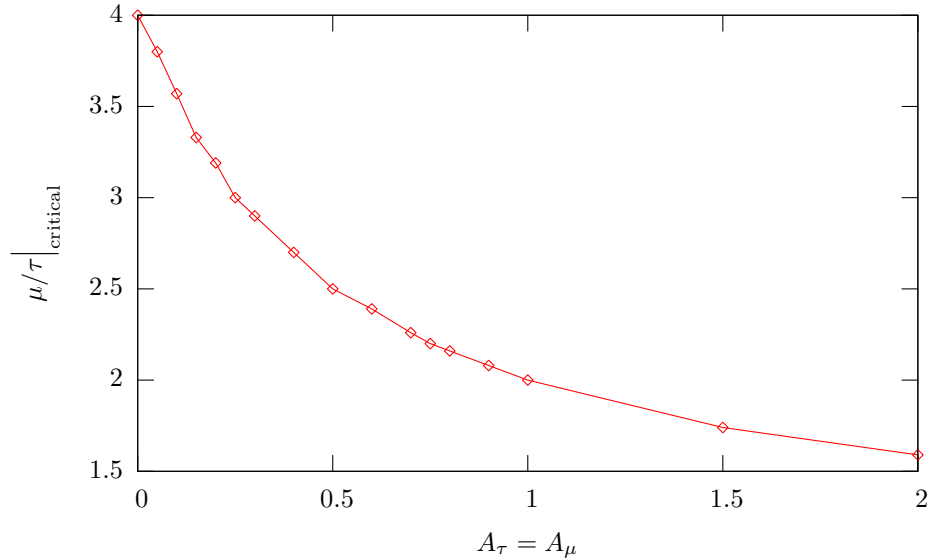}
  \caption{\label{fig10} For the lattice refinement models $\delta_\mu
    = \mu^{-A_\mu}$ and $\delta_\tau= \tau^{-A_\tau}$, the condition
    of stability has been found. Here we plot $\mu/\tau$ along the
    critical line, i.e., the line in which the system is just becoming
    unstable, as a function of the parameter $A$.}
 \end{center}
\end{figure}

By repeating this over a range of $\mu$-positions for the
perturbation, we are able to empirically confirm the stability
condition $\mu < 2 \tau$. We use this method to investigate how the
lattice refinement model alters the stability properties of the
system. In particular, we find that for $\delta_\mu = \mu^{-A}$ and
$\delta_\tau = \tau^{-A}$, the stability condition ranges between $\mu
< 4\tau$ for the $A=0$ case (constant lattice) and $\mu < 1.58\tau$
for the case of $A = 2.0$.  Figure \ref{fig10} shows the cases in
between. In all cases, the second-order correction terms are at
least an order of magnitude lower than the wave-functions.

It is worth noting that for the constant lattice case, one does not
have to perform any interpolations to evaluate the
wave-function. Thus, one does not require the wave-function to be {\sl
  pre-classical} and large changes between successive lattice points
are perfectly acceptable. However, such wave-functions clearly do not
have a semi-classical limit\footnote{When plotting these results, only
  values at the lattice points should be used, since the system is
  inherently discrete, however to guide the eye we have plotted the
  results with lines.}. This is not true for the varying lattice
cases, where for the interpolation to have a significant meaning, one
must require that the wave-function be {\sl pre-classical} (so that
the derivatives in the Taylor expansion are small). We ensure that the
stability condition holds by taking a small perturbation ($\approx
10^{-6}$).  This is demonstrated well by the fact that for the case of
$A=-1$, our results confirm the analytic considerations
of Ref.~\cite{Bojowald:2007ra}.

In Ref.~\cite{Bojowald:2007ra} a second lattice refinement model was
shown to be unconditionally stable, under certain
circumstances. Specifically, for $\delta_\mu\left( \mu,\tau\right) =
\mu_0 \sqrt{\tau} \mu^{-1}$ and $\delta_\tau \left(\mu,\tau\right) =
\tau_0 \tau^{-1/2}$, assuming the solutions do not change
significantly on the scale of the step-sizes, the difference equation
was found to be stable~\cite{Bojowald:2007ra}. We find that this is
indeed true, however as soon as the wave-functions fail to be {\sl
  pre-classical}, i.e., as soon as there is a significant variation
between lattice points (of the order of a few percent), they become
unstable.
\begin{figure}
 \begin{center}
  \includegraphics{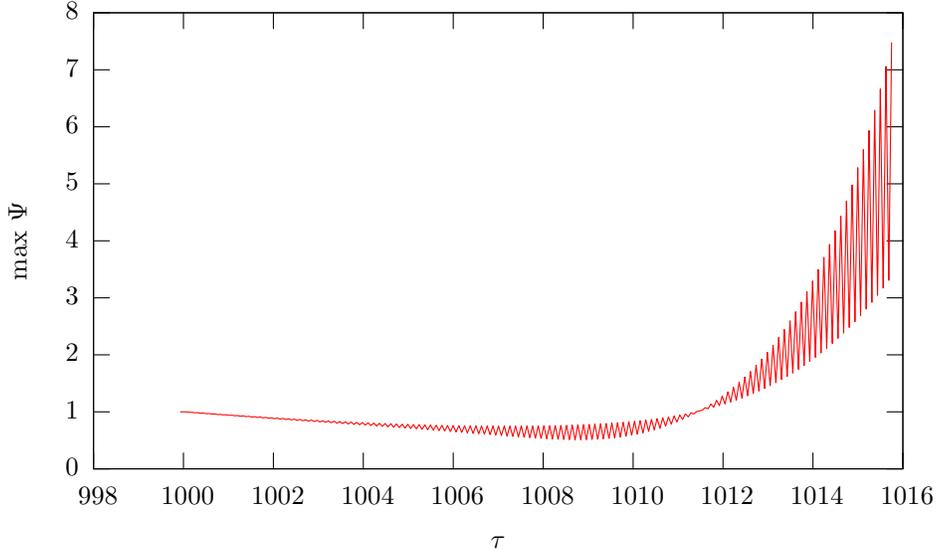}
  \caption{\label{fig12} The maximum amplitude for a Gaussian
    wave-packet centred on $\mu = 1060$ of width $\sigma = 5.25$,
    using the lattice refinement model $\delta_\mu = \mu_0 \sqrt{\tau}
    \mu^{-1}$ and $\delta_\tau = \tau_0 \tau^{-1/2}$. For small $\tau$
    the solution does not vary significantly on scales of the order of
    the lattice size and hence the solution is stable, however as soon
    as the solution begins to vary this stability is lost.}
 \end{center}
\end{figure}
In Fig.~\ref{fig12} we evaluate a Gaussian (centred on $\mu = 1060$
of width $\sigma = 5.25$) for this lattice refinement model. This
solution is not entirely {\sl pre-classical}, since there is a
variation of $\approx 0.1\%$ between the wave-function evaluated on
successive $\tau$ lattice points. This variation grows and when it
reaches the order of a few percent the system becomes unstable. The
rate at which small initial variations grow, and hence the rate at
which the instability becomes apparent, depends on the $\mu$- and
$\tau$-coordinates, growing faster for large $\mu$ and small $\tau$.

Nevertheless, these instabilities do not represent any significant
problem for loop quantum gravity approaches to black holes. It is
already know that solutions must be {\sl pre-classical} in order to
have a well-defined continuum limit (i.e., in order for $\lim_{\mu
  \rightarrow \infty, \tau \rightarrow \infty}\Psi_{\mu,\tau} =
\Psi\left(\mu,\tau\right)$ to be valid). What is important is that
this continuum limit is always stable, which is indeed the case for the
$\delta_\mu = \mu_0 \sqrt{\tau} \mu^{-1}$, $\delta_\tau = \tau_0
\tau^{-1/2}$ lattice refinement model, but not for models of the form
$\delta_\mu\left(\mu,\tau\right)= \mu_0\mu^{-A}$, $\delta_\tau =
\left(\mu, \tau\right) = \tau_0\tau^{-A}$.

The presence of these instabilities leads to difficulties in the
numerical implementation of the method described above. In particular,
if one wants to calculate the wave-function for large values of
$\tau$, then one needs to start with large range of $\mu$, to ensure
that enough initial data is known ({\sl see}, Fig.~\ref{fig14}). This
can mean that the system is unstable at the large $\mu$-side of the
lattice, for small initial $\tau$. 

\begin{figure}
 \begin{center}
  \includegraphics[scale=0.75]{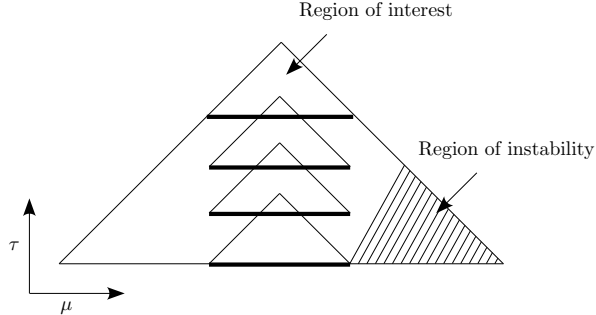}
  \caption{\label{fig11} The instability present in the difference
    equation occurs for large $\mu$ and small $\tau$. If one is
    interested in reaching the value of the wave-function at large
    $\tau$, one needs to have a large initial range of $\mu$ to have
    sufficient data. This can result in a portion of the domain being
    in the unstable region, which makes the code very sensitive to
    numerical inaccuracies. To avoid this, one can simply use a
    restricted $\mu$-range and evaluate the wave-function as far as
    possible. This wave-function can then be used to create the
    initial conditions for a subsequent evaluation, provided the
    wave-function has not reached the edge of the lattice.}
 \end{center}
\end{figure}

In practise, the value of the wave-function would usually be zero in
this region, since one is typically interested in how an initial
wave-packet evolves. Any non-zero component of the wave-packet in this
unstable region would have met the right-hand diagonal edge of the
lattice before the high $\tau$-region of interest is reached.  However
rounding errors can result in a non-zero perturbation, which will grow
exponentially due to the inherent instability of the difference
equation. Such difficulties can be overcome, by evaluating the
wave-function over a small region of $\tau$. Since we now only require
a smaller region of $\mu$ for this evaluation, the unstable region can
be avoided. To reach the high $\tau$-region of interest, the system
can be re-set using the out-putted wave-function added to a larger
range of $\mu$, see Fig.~(\ref{fig11}).  Alternatively, the additional
initial data necessary to reach the large value of $\tau$ could be
included at the evaluation of each $\mu$-line, however this requires
an additional three sets of $\tau$-data, which effectively doubles the
amount of initial data required.

\section{Propagation of the wave-function}
\begin{figure}
 \begin{center}
  \includegraphics[scale=1.0]{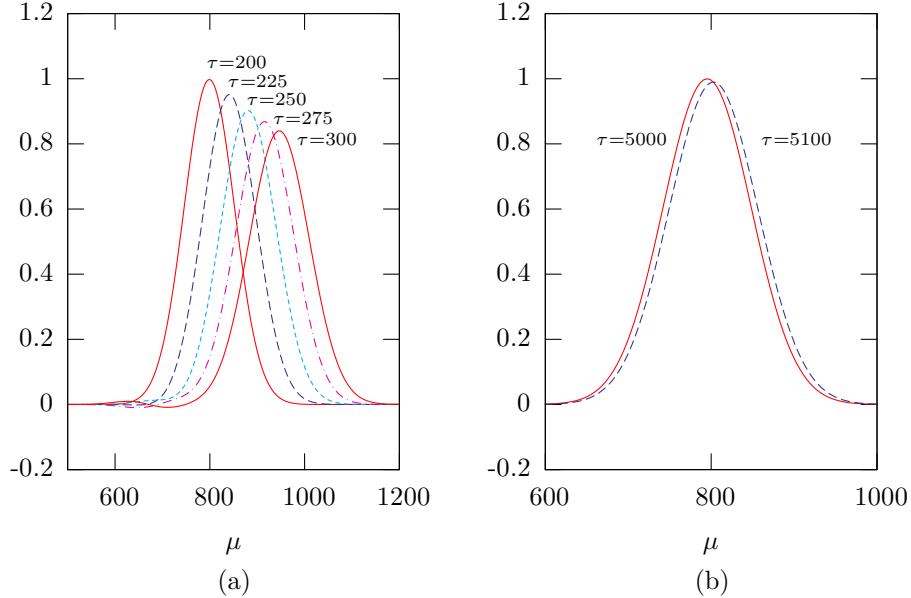}
  \caption{\label{fig13} (a) A Gaussian centred on
    $\left(\mu,\tau\right) = \left( 800,200\right)$ {\it twists} to larger
    values of $\mu$ as $\tau$ is increased. (b) As we move into the
    continuum limit this effect disappears. In this case the Gaussian
    is centred on $\left(\mu,\tau\right) =\left(800,5000\right)$.
    Here the constant lattice case is shown however this effect persists
    when lattice refinement is modelled.}
 \end{center}
\end{figure}

\begin{figure}
 \begin{center}
  \includegraphics[scale=0.75]{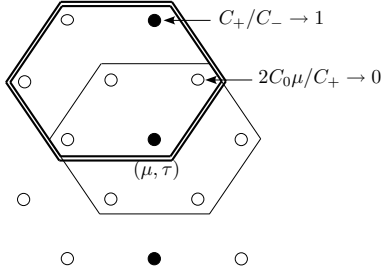}
  \caption{\label{fig15}  By considering all the lattice points on the initial
$\mu$ row except one being zero, the resulting wave-function can be
calculated. When the discreteness cannot be ignored, there is a net {\it motion}
towards larger values of $\mu$ $(\Theta)$. The lower hexagon shows the lattice
points needed to evaluate the first step of the wave-function, whilst the
upper hexagon highlights the points needed for the second step. The solid circles
represent lattice points at which the wave-function is unity whilst at the open
circles the wave-function is zero. From these it can be shown that the {\it twist}
induced on the wave-function disappears on large-scales.}
 \end{center}
\end{figure}

In the case of a constant lattice, an initially centred Gaussian will
move to larger $\mu$, as $\tau$ is increased ({\sl see},
Fig.~\ref{fig13}). [In terms of more usual coordinates, the angle
$\Theta$ increases as the radius increases.] The reason for this can
be seen simply by considering the original difference equation for the
simple case of a lattice point on which $\Psi_{\mu,\tau}=1$ and all
other known $\Psi$-values are zero. Schematically, it is given by the
lower hexagon in Fig.~\ref{fig15}, centred around the non-zero
wave-function $\Psi_{\mu,\tau}$.  Then, Eq.~(\ref{eq:vary2D}) with
$\gamma=0$, implies
$$\Psi_{\mu+2\delta_\mu,\tau+2\delta_\tau}=2C_0\mu/C_{+}~.$$ For $\tau
\gg \delta_\tau$, this goes to zero. To see that this implies no
motion of the wave-function, consider the next (upper) hexagon in
Fig.~\ref{fig15}, centred around the lattice point
$\Psi_{\mu-2\delta_\mu,\tau+2\delta_\tau}$. If we `update' the coordinates
so that this central point is again called $\left(\mu,\tau\right)$, then,
in these coordinates, only the lower right hand point in this (upper)
hexagon is non-zero,
$$\Psi_{\mu+2\delta_\mu,\tau-2\delta_\tau} = 1~.$$
In this case, Eq.~(\ref{eq:vary2D}) gives
$$\Psi_{\mu+2\delta_\mu, \tau+2\delta_\tau} = C_{-}/C_{+}~,$$ 
which tends to unity for $\tau \gg \delta_\tau$. Thus, we find that
for large $\tau$ a value at one lattice point $\left(\mu,\tau\right)$,
moves to one with the same $\mu$-coordinate,
$\left(\mu,\tau+4\delta_\tau\right)$, i.e., there is no
motion.

 However, when we are in a region in which the lattice
discreteness is important, this is no longer true, as
$2C_0\mu/C_{+}\neq 0$ and the value at one lattice point introduces a
non-zero component to the value at a lattice point with larger $\mu$
coordinate, i.e., the wave-function moves to larger $\mu$. This implies
the existence of some induced rotation on the wave-function due to the
underlying discreteness of the space-time.

If we include lattice refinement, then the same effect occurs. Once
again, there is no motion for $\tau \gg \delta_\tau$ and in the case
of lattice refinement this requirement is reached for lower $\tau$,
since $\delta_\tau$ reduces as $\tau$ increases. Thus, we expect that
the effect will disappear quicker than in the constant lattice case,
because the lattice refinement brings us into the continuum limit
faster. However once again, as the wave-function moves into a region
in which the discreteness of the lattice is important, a motion will
be induced.

\section{Conclusions}
Here, we have developed a simple and intuitive prescription for
evaluating two-dimensional wave-functions to a well-controlled level
of accuracy, for arbitrary lattice refinement models. We focused on
black-hole interiors, however the method clearly extends to
anisotropic Bianchi models and other systems with anisotropic
symmetries. 

We have shown how the stability conditions on the Hamiltonian
constraint can be investigated using this numerical method and
extended the range of lattice refinement models for which the
stability criterion are known. 

We have also examined and explained the existence of a {\it twist} in
the wave-functions, due to the underlying discreteness of the theory; a
feature that warrants further study, particularly in relation to its
effect in microscopic black holes.

\section{Acknowledgments}
This work is patially supported by the European Union through the
Marie Curie REsearch and Training Network {\sl Universeet}
(MRTN-CT-2006-035863).


\begin{thebibliography}{10}
\bibitem{rovelli2004}
C.~Rovelli, {\sl Quantum Gravity} (Cambridge University Press, Cambridge, 2004).

\bibitem{Ashtekar:2003hd}
  A.~Ashtekar, M.~Bojowald and J.~Lewandowski,
  Adv.\ Theor.\ Math.\ Phys.\  {\bf 7} (2003) 233
  [arXiv:gr-qc/0304074].

\bibitem{Bojowald:2002gz}
  M.~Bojowald,
  Class.\ Quant.\ Grav.\  {\bf 19} (2002) 2717
  [arXiv:gr-qc/0202077].

\bibitem{Rosen:2006bga}
  J.~Rosen, J.~H.~Jung and G.~Khanna,
  Class.\ Quant.\ Grav.\  {\bf 23} (2006) 7075
  [arXiv:gr-qc/0607044].

\bibitem{Bojowald:2007ra}
  M.~Bojowald, D.~Cartin and G.~Khanna,
  Phys.\ Rev.\  D {\bf 76} (2007) 064018
  [arXiv:0704.1137 [gr-qc]].

\bibitem{Nelson:2007um}
  W.~Nelson and M.~Sakellariadou,
  Phys.\ Rev.\  D {\bf 76} (2007) 104003
  [arXiv:0707.0588 [gr-qc]].

\bibitem{Sabharwal:2007xy}
  S.~Sabharwal and G.~Khanna,
  ``Numerical solutions to lattice-refined models in loop quantum cosmology,''
  arXiv:0711.2086 [gr-qc].

\bibitem{Ashtekar:2006wn}
  A.~Ashtekar, T.~Pawlowski and P.~Singh,
  Phys.\ Rev.\  D {\bf 74} (2006) 08400
  [arXiv:gr-qc/0607039].

\bibitem{Vandersloot:2005kh}
  K.~Vandersloot,
  Phys.\ Rev.\  D {\bf 71} (2005) 103506
  [arXiv:gr-qc/0502082].

\bibitem{Ashtekar:2000eq}
  A.~Ashtekar, J.~C.~Baez and K.~Krasnov,
  Adv.\ Theor.\ Math.\ Phys.\  {\bf 4} (2000) 1
  [arXiv:gr-qc/0005126].

\bibitem{Ashtekar:2003zx}
  A.~Ashtekar and A.~Corichi,
  Class.\ Quant.\ Grav.\  {\bf 20} (2003) 4473
  [arXiv:gr-qc/0305082].

\bibitem{Nelson:2007wj}
  W.~Nelson and M.~Sakellariadou,
  Phys.\ Rev.\  D {\bf 76} (2007) 044015
  [arXiv:0706.0179 [gr-qc]].

\bibitem{Bojowald:2002ny}
  M.~Bojowald,
  Class.\ Quant.\ Grav.\  {\bf 19} (2002) 5113
  [arXiv:gr-qc/0206053].

\bibitem{Ashtekar:2004eh}
  A.~Ashtekar and J.~Lewandowski,
  Class.\ Quant.\ Grav.\  {\bf 21} (2004) R53
  [arXiv:gr-qc/0404018].
\end{thebibliography}
\end{document}